\documentclass[prl,superscriptaddress,showkeys,showpacs,10pt,tightenlines,floatfix,twocolumn]{revtex4}
\usepackage[utf8]{inputenc}
\usepackage{amsmath}
\usepackage{amsfonts}
\usepackage{amssymb}
\usepackage{graphicx}
\usepackage{color}
\usepackage[caption=false]{subfig}

\newcommand{\ie}{{\it i.e. }}
\newcommand{\br}{\mathbf{r}}

\newcommand{\revision}[1]{#1}

\graphicspath{{./images/}}
	
\begin{document}
\title{Thermally driven order-disorder transition in two-dimensional soft cellular systems}
\author{Marc Durand and Julien Heu}
\affiliation{Mati\`{e}re et Syst\`{e}mes Complexes (MSC), 10 rue Alice Domon et L\'{e}onie Duquet, 75205 Paris Cedex 13, France.}
 \altaffiliation{UMR 7057 CNRS \& Universit\'{e} Paris Diderot}
\email{marc.durand@univ-paris-diderot.fr} 
\date{\today}

\begin{abstract}
Many systems, including biological tissues and foams, are made of highly packed units having high deformability but low compressibility. At two dimensions, these systems offer natural tesselations of plane with fixed density, in which transitions from ordered to disordered patterns are often observed, in both directions.
Using a modified
Cellular Potts Model algorithm that allows rapid thermalization of extensive systems, we numerically explore the order-disorder transition of monodisperse, two-dimensional cellular systems driven by thermal agitation. 
We show that the transition follows most of the predictions of Kosterlitz-Thouless-Halperin-Nelson-Young (KTHNY) theory developed for melting of 2D solids, extending the validity of this theory to systems with many-body interactions.
In particular, we show the existence of an intermediate hexatic phase, which preserves the orientational order of the regular hexagonal tiling, but looses its positional order.
In addition to shedding light on the structural changes observed in experimental systems, our study shows that soft cellular systems offer macroscopic systems in which KTHNY melting scenario can be explored, in the continuation of Bragg's experiments on bubble rafts.

\end{abstract}
\maketitle

\begin{figure}[h]
	\begin{center}
		\includegraphics[width=\columnwidth]{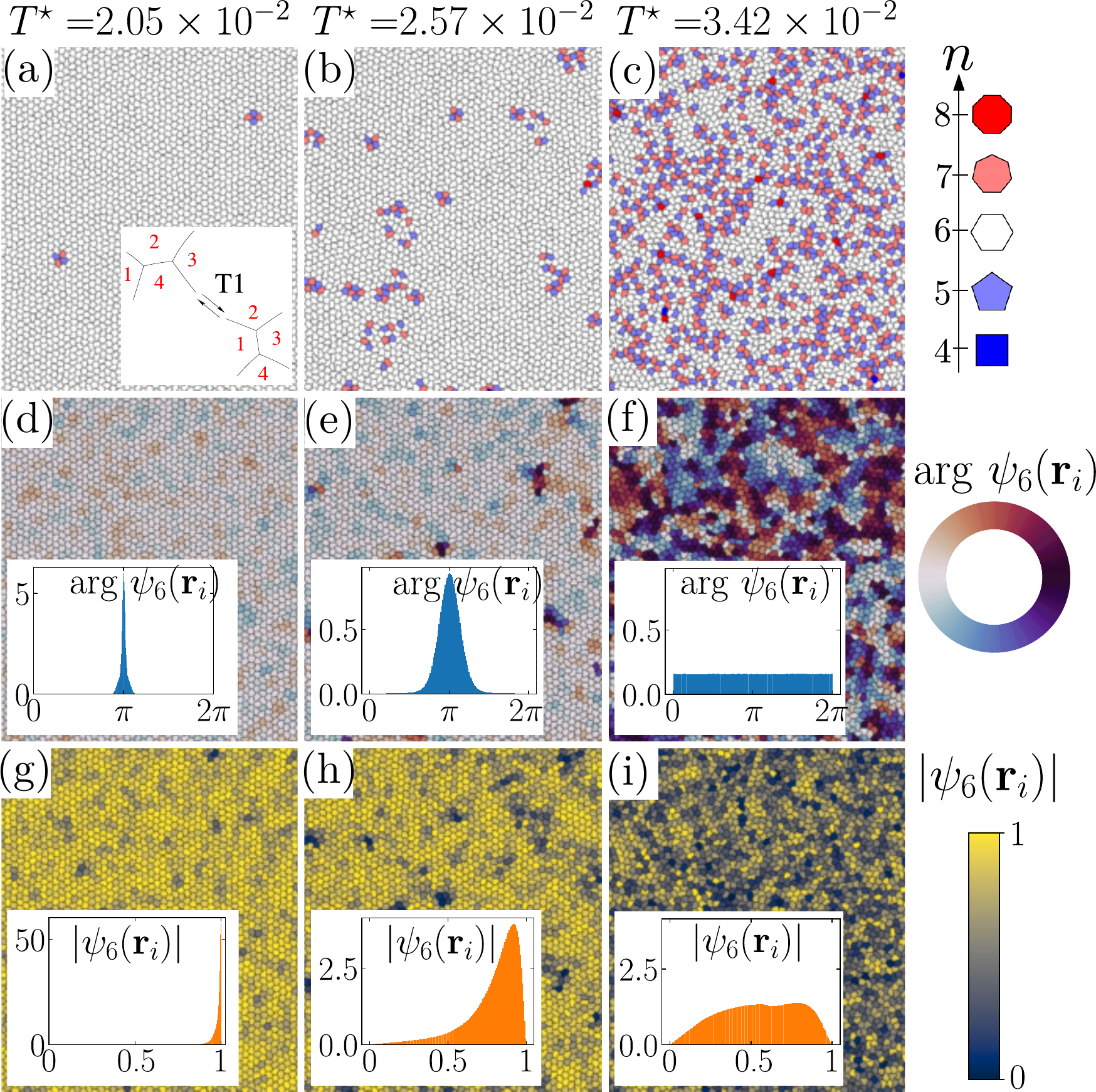}
	\end{center}
	\caption{Evolution of defect populations and orientational order with temperatures. Only a small portion of the system is shown. Columns from left to right: solid, hexatic, and liquid phases. 
		Top row (a--c): population of topological defects. Inset: sketch of a T1 event: in an hexagonal lattice, this topological change corresponds to the creation of a bounded pair of dislocations.
		Middle row (d--f): phase of the local orientational parameter order $\mathrm{arg}~\psi_6 (\mathbf{r}_i)$. Insets: normalized histograms of $\mathrm{arg}~\psi_6 (\mathbf{r}_i)$, calculated over the whole simulation time.
		Bottom row (g--i): norm of the local orientational parameter order $\vert \psi_6 (\mathbf{r}_i)\vert$. Insets: normalized histograms $\vert \psi_6 (\mathbf{r}_i)\vert$, calculated over the whole simulation time.\label{fig:panel}}	
\end{figure}

Foams, emulsions, and confluent biological tissues are examples of \textit{Soft Cellular Systems} (SCS): 
They are constituted of highly deformable -- yet almost incompressible  -- units (bubbles, drops, cells,\dots), interacting through attractive adhesive interactions and soft steric repulsions. When
highly compacted, they tile the available space (3D) or plane (2D) perfectly, \ie without gaps or overlaps.
Interface energy is key to the cohesion and the rigidity of these systems,
sometimes constituted solely of fluids.
SCS have rough (mechanical) energy landscapes with many local minima: 
for a given number of units (with prescribed sizes), many different tilings are possible. In some cases the tiling is ordered, consisting predominantly of hexagons (in 2D), and in others it is disordered and includes a distribution of topological defects (polygons with $n \neq 6$ sides). On the timescales considered here, the transition from one local minimum to another is achieved through a succession of elementary structural rearrangements, called T1 events (see inset in Fig. \ref{fig:panel}a), which preserve the integrity and size of the cellular units \cite{Footnote}.

Over the last years,
special attention has been given to the glass transition in these systems \cite{Aste_1999,Davison_2000,Angelini_2011,Bi_2015,Bi_2016}, \textit{i.e.} in the transition from a disordered, solid phase to a disordered, liquid phase.
In many situations, though -- and especially during morphogenetic movements -- we observe a transition between ordered and disordered patterns, in both directions \cite{Quilliet_2008,Durand_statistical_2011, Durand_statistical_2014,Zallen_2004,Classen_2005,Hocevar_2009}. Such order-disorder transitions received little attention from a theoretical point of view \cite{Staple_2010}.
In systems that are monodisperse in size, the structural disorder is of topological origin only: for 2D SCS, hexagonal tiling is the only monodisperse regular tiling, and structural disorder arises from the presence of non-hexagonal cells, or \emph{topological defects}, that are generated by T1 events. Structural disorder strongly affects the mechanical properties of a tissue (or any other SCS), 
 and is also essential for its function. 
 Eventually, the study of the pattern of SCS or its fluctuations \cite{Sussman_2018, Fodor_2018} can teach us about its mechanical properties and provide a tool for diagnosis.

In the present letter, we numerically 
investigate the order-disorder phase transition in monodisperse 2D cellular systems, using a modified
Cellular Potts Model algorithm that allows thermalization of large systems. 
Order-disorder transition in SCS is usually driven by a non-thermal source of energy: in a confluent tissue, T1s are consequences of cell activity powered with chemical energy (ATP). In passive SCS such as foams or emulsions, T1s are induced by the injection of  mechanical energy through the application of some mechanical stress. 
Here, we modelize these out-of-equilibrium dynamics by an effective simulation temperature \cite{Sussman_2018}, and investigate the phase transition driven by this effective thermal agitation. 
%
Thermally driven transition will offer a benchmark system over which we will build up when comparing with actively driven transitions. 
%
Since the pioneering Bragg's experiments \cite{Bragg_1947, Bragg_1949}, foams and bubble rafts have been recognized as macroscopic model systems for studying the geometry, the dynamics, and the deformation behavior of atomic or molecular materials.
We show in the present study that foams also provide a macroscopic model for studying melting of two-dimensional materials.

Order-disorder transition in 2D SCS is reminiscent of defect-mediated theories for melting of two-dimensional solids.
%
%
In these theories, the phase transition is described in terms of topological defects in the Voronoi partition associated with the lattice of the solids. 
Defect-mediated melting theories have been tested, experimentally and numerically, on a large variety of systems, \revision{including Lennard-Jones systems \cite{Chen_1995}, colloidal particles \cite{vonGrunberg_2007, Marcus_1996}, magnetic beads \cite{Schockmel_2013}, and disks with either hard-core or soft potentials \cite{Kapfer_2015}.}
SCS yet differ in different aspects from all these classes of systems: firstly, the partition of 2D space is a physical partition, not a mathematical construction like Voronoi tesselation. Hence, energy and partition are directly related. Secondly, due to their unique high deformability--low compressibility feature, interactions between the cellular units are not pairwise additive \cite{Bi_2016,Hohler_2017,Ginot_2019}. Many-body interactions are known to affect the mechanical properties of SCS \cite{Hohler_2017}. They also make the phase transition scenario uncertain.

The most popular defect-mediated melting scenario is provided by the Kosterlitz-Thouless-Halperin-Nelson-Young (KTHNY) theory \cite{Strandburg_1988, Glaser_1993, vonGrunberg_2007}, 
which predicts two-step melting,
from the crystal to a intermediate hexatic phase and then from the hexatic
to a liquid phase. The two transitions are associated with the disappearance of translational and orientational orders, successively. The intermediate hexatic phase has short-range translational order but quasi-long-range orientational order. The dissociation of bound dislocation pairs into free dislocations drives the solid into the hexatic phase, while the unbinding of dislocations into isolated disclinations drives the hexatic to liquid transition. 
Other melting scenarios are however possible. Those based on proliferation of vacancies or interstitials \cite{Glaser_1993} are irrelevant for SCS, in which such defects cannot take place. 
%
Another popular scenario argues that for systems with core
energy of dislocations not too large compared to $k_B T$, melting is caused by nucleation and proliferation of grain-boundaries, preempting the hexatic phase \cite{Saito_1982, Chui_1983}. 

Our simulations are based on the Cellular Potts Model (CPM), which is widely used for simulating
cellular systems in various fields of physics or biology, such as coarsening and mechanics of foams \cite{Glazier1990,Jiang_1999}, tissue morphogenesis \cite{Hirashima_2017}, cell sorting \cite{Graner_1992} and collective cell motion in epithelial tissues \cite{Szabo_2010, Kabla_2012}.
%
The CPM is a lattice based modeling technique:
each cell is represented as a subset of lattice sites sharing the same cell ID (analogical to spins in Potts model).
Cellular domains can adopt any shape on the lattice.
The CPM is then particularly suited to simulate thermal fluctuations of cellular systems, as it reproduces realistically the fluctuations of interface locations, even for wavelength at subcellular scale. Furthermore, its extension to three dimensions is straightforward.
The system evolves using a recently modified Metropolis algorithm that preserves the integrity of the cellular domains and satisfies the detailed balance equation \cite{Durand_2016}, ensuring that the probability distribution of visited states converges to the Boltzmann distribution. This algorithm has also been proved to be more efficient that the standard algorithm used in CPM for a same simulation temperature $T$ \cite{Durand_2016}, allowing us to simulate much larger systems.

Mechanical energy of monodisperse SCS is modeled by the discretized version of the following Hamiltonian:

\begin{equation}
\mathcal{H}=\gamma\sum_{\substack{ \langle i,j \rangle}}L_{ij}+\dfrac{B}{2A_0}\sum_{\substack{ i}}\left(A_i-A_0\right)^2.
\label{Hamiltonian}
\end{equation}
%
The first term in Eq. \ref{Hamiltonian} accounts for interfacial effects: the sum is carried over neighboring cells $\langle i,j \rangle$ and $L_{ij}$ is the boundary length between cells $i$ and $j$. The second term accounts for an effective area elasticity which results from a combination of three-dimensional
cell incompressibility and cell bulk elasticity. $B$ is the effective 2D bulk modulus, $A_i$ is the actual area of cell $i$, and $A_0$ the preferred cell area. 
%
This simplified version of the standard Hamiltonian used for cellular systems \cite{Staple_2010,Bi_2015,Bi_2016} corresponds to the situation of a foam or an epithelium with inflated shapes \cite{Suppl_Mat,Staple_2010}. 

\revision{We emphasize that Hamiltonian (\ref{Hamiltonian}) cannot be expressed as a sum of pairwise interactions:
 because cellular units have high deformability but low compressibility, they adjust their shape rather than their size when brought in contact. In a confluent system, any modification of the interface between cells $i$ and $j$ will imply a displacement of the two cell centers, but also of the other neighboring cells in order to tile the plane perfectly, while preserving cell areas. As a consequence, contact length $L_{ij}$ does not only depend on the relative position of the two adjacent cells  $i$ and $j$, but also on the positions of other neighboring cells  \cite{Bi_2016,Hohler_2017,Ginot_2019}.}

The relative importance of thermal, interfacial and bulk energies are quantified with two dimensionless parameters: the reduced temperature $T^\star = T/(\gamma a)$ and the reduced compressibility $\xi=\gamma/\left( B a \right)$, where $a$ is the equivalent hexagonal lattice step: $a=(2A_0/\sqrt{3})^{1/2}$. $\xi$ must be low enough to reproduce accurately real systems, but \revision{must remain} finite to allow for some area fluctuations required by the Metropolis-like algorithm. 

We have performed extensive numerical simulations of systems of $N = 200^2=40,000$ cells in a rectangular lattice of $1400\times 1616$ sites under periodic conditions (then $A_0=56.56$ pixels$^2$). Periodic boundary conditions allow global translations of the lattice, but forbid global rotations.
The lattice aspect ratio approximates the target value of $2/ \sqrt 3$, corresponding to the aspect ratio of a perfect hexagonal lattice, with error less than $4.10^{-4}\%$. For all simulation runs, we start with the same ordered hexagonal tiling (honeycomb), and wait for equilibration before recording data (see \cite{Suppl_Mat} for details on estimation of equilibration time). 
In all our simulations, we choose $\gamma=180$, $B=200$ and so $\xi \simeq 0.11$.

Figs. \ref{fig:panel}(a--c) show the equilibrated system at three different temperatures, with non-hexagonal cells that are color-coded. At low temperature (\ref{fig:panel}a), the only topological defects are bound dislocation pairs. At intermediate temperature (\ref{fig:panel}b), paired and single dislocations coexist. At high temperature (\ref{fig:panel}c), most of topological defects are assembled into aggregates. 

\begin{figure}[h]
	\centering
	\subfloat[]{	
	\includegraphics[width=0.5\columnwidth]{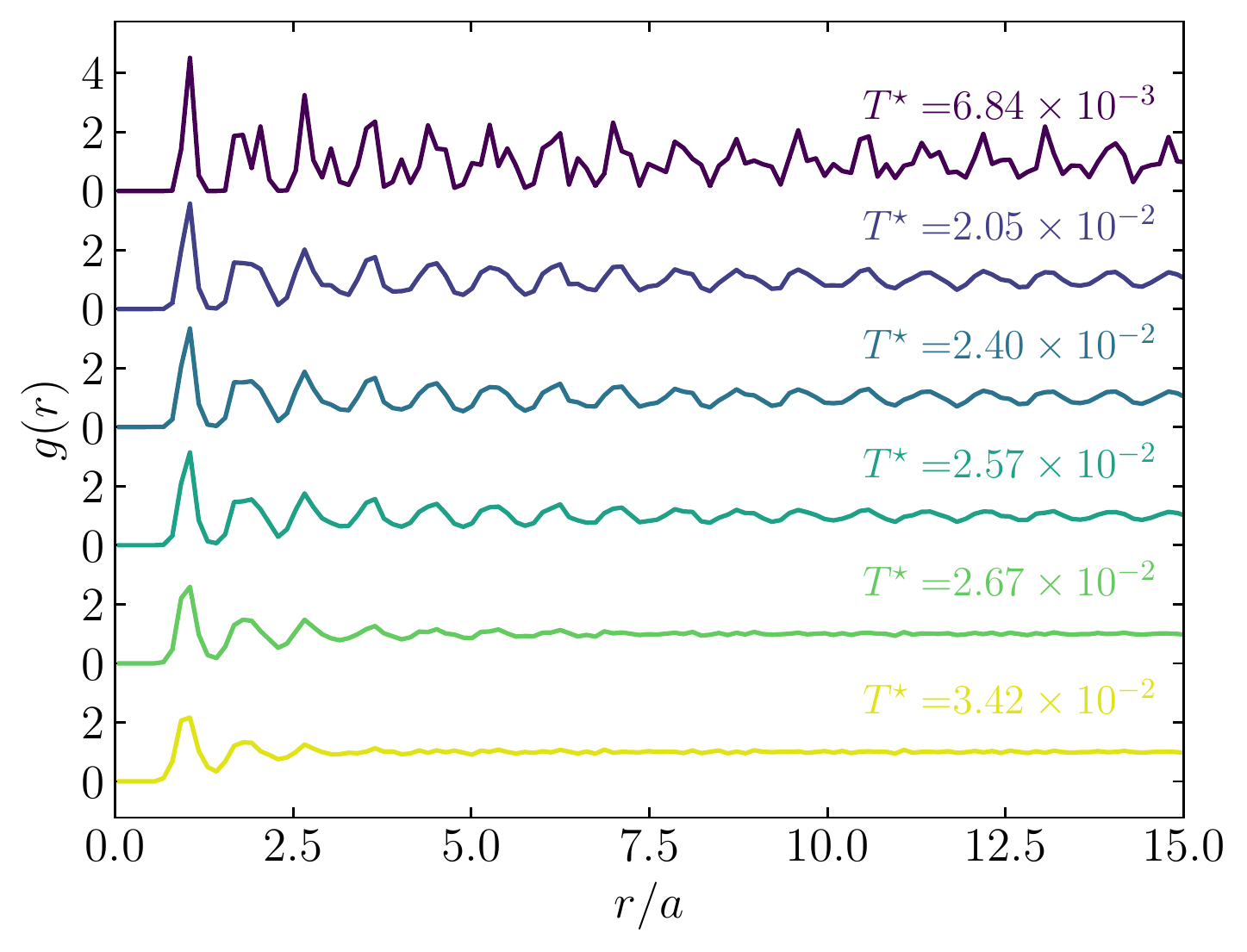}
	\label{fig:plots_gr}
	}
	\subfloat[]{	
	\includegraphics[width=0.5\columnwidth]{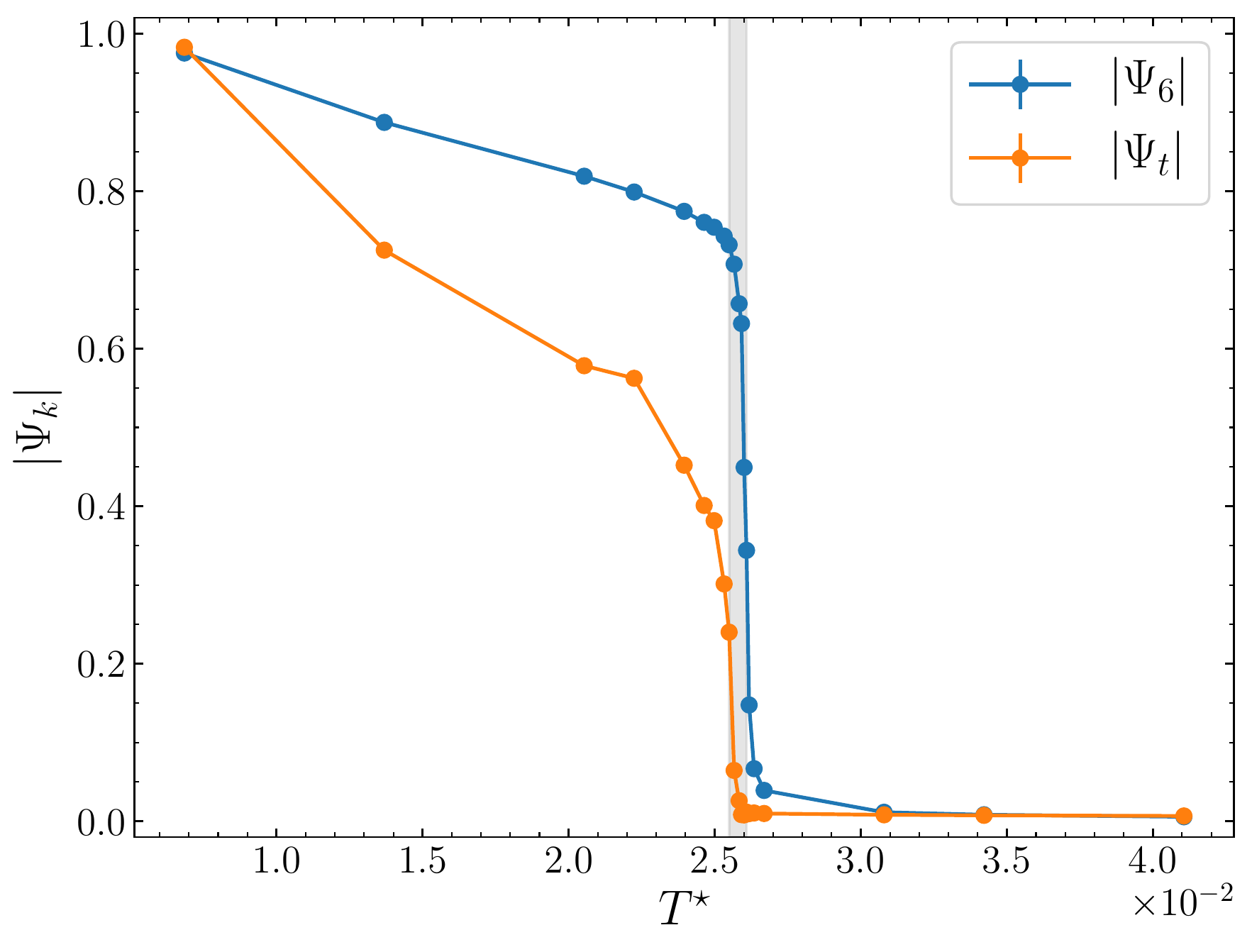}
	\label{fig:abs_para}
	}
	\caption{(a) Pair correlation function $g(r)$ at different reduced temperatures. Plots are evenly shifted vertically for clarity. Peaks are smeared out as temperature increases,
	indicating the loss of translational order.
	(b) Variation of $\vert \Psi_{t} \vert$ and $\vert \Psi_6 \vert$ with temperature. The drop of $\vert \Psi_{t} \vert$ slighlty precedes $\vert \Psi_6 \vert$ drop, indicating two distinct transitions. 
	The shaded area indicates the existence domain of the hexatic phase as determined from the peaks of the two susceptibilities (see Fig. \ref{fig:plot_susceptibilities}). Standard errors of the mean are smaller than the size of the markers. 
	}
\end{figure}
%
To characterize the static structure of the cellular system occurring during 2D melting, we first calculate the pair correlation function $g(r)$, defined as
\begin{equation}
\revision{
	g(r)=\dfrac{1}{N}\left\langle \sum_{j\neq k} \delta \left( r-\vert \br_j-\br_k \vert \right) \right\rangle,
}
\end{equation}
\revision{where $\mathbf{r}_j$ and $\mathbf{r}_k$ are the geometric center positions of cells $j$ and $k$, respectively.} The angular brackets denote an average over central cell $k$. This correlation function gives the probability to find two
cells separated by a distance $r$. Fig. \ref{fig:plots_gr} shows that the peaks of $g(r)$ get broader and shorter as temperature increases, 
indicating that the system melts from an ordered crystal to a disordered liquid.
%
%


To get more insights into the structural change during the melting process, we focus on the two order parameters which characterize the translational and orientational symmetries of the system, respectively. The global translational order parameter is defined by
\begin{equation}
\Psi_{t}=\dfrac{1}{N}\sum_{i=1}^N \psi_{t}(\mathbf{r}_i)
\end{equation}
where $\psi_{t}(\mathbf{r})=\exp{(i \mathbf{G} \cdot \mathbf{r})}$  is the local translational
order parameter for cell $i$ at position $\br_i$, and $\mathbf{G}$ is a primary reciprocal lattice vector.  
The global orientational order parameter is given by
\begin{equation}
\Psi_6=\dfrac{1}{N}\sum_{i=1}^N \psi_{6}(\mathbf{r}_i),
\end{equation}
where $\psi_6 (\mathbf{r}_i) = \frac{1}{z_i}\sum_{j\in z_i}\exp{(i6\theta_{ij})}$ is the local orientational parameter order,
$z_i$ is the number of neighbors of cell $i$, and $\theta_{ij}$ is the angle of the bond between centers of cell $i$ and $j$ relative to a fixed reference axis. The magnitude of the local orientational order, $\vert \psi_6(\br_i) \vert$ ranges from $0$ to $1$ and measures the degree to which the cell's neighbourhood resembles a hexagonal
crystal, while its phase $\mathrm{arg}~\psi_6(\br_i)$ indicates the local lattice director.

 Spatial distributions of $\mathrm{arg}~\psi_6(\br_i)$ and $\vert \psi_6(\br_i) \vert$ are shown in Figs. \ref{fig:panel}(d--i) at three different temperatures. Spatial heterogeneities increase with temperature.
%
Normalized histograms of $\vert \psi_6(\br_i) \vert$ and $\mathrm{arg}~\psi_6(\br_i)$ are shown in the corresponding insets. Both quantities are clearly peaked in the solid and intermediate phases, revealing that the orientational order is preserved in these two phases. Peaks are more spread out in the intermediate phase, indicating a quasi-long orientational order in this phase, as expected for the hexatic phase predicted by the KTHNY theory. In the liquid phase, distributions are flat and the orientational order is lost.

%
Fig. \ref{fig:abs_para} shows the variation of the global order parameters $\vert \Psi_{t} \vert$ and $\vert \Psi_6 \vert$ with temperature.
Both curves present similar tendencies: 
%
at low temperature, both order parameters decay slowly with $T$. The decay is more pronounced for $\vert \Psi_t \vert$, suggesting a \emph{quasi} long-range orientational order in the solid phase. 
%
%
Then, both curves drop abruptly to $\simeq 0$. $\vert\Psi_t\vert$ falls slightly before $\vert\Psi_6\vert$, suggesting the existence of two distinct phase transitions, and subsequently the existence of an intermediate hexatic phase, in agreement with KTHNY scenario.




To further characterize the transition, we now analyze the correlation functions of the two local order parameters, 
for which specific behaviors are expected within the KTHNY theory \cite{Strandburg_1988, Glaser_1993, vonGrunberg_2007}. They are defined as

\begin{equation}
C_{k}(r) =\vert \langle \psi_{k}^*(\mathbf{r}'+\mathbf{r}) \psi_{k}(\mathbf{r}') \rangle_{r'} \vert
\end{equation}
with $k \equiv 6,t$.
\begin{figure}[h]
	\centering
	\subfloat[]{	
		\includegraphics[width=0.5\columnwidth]{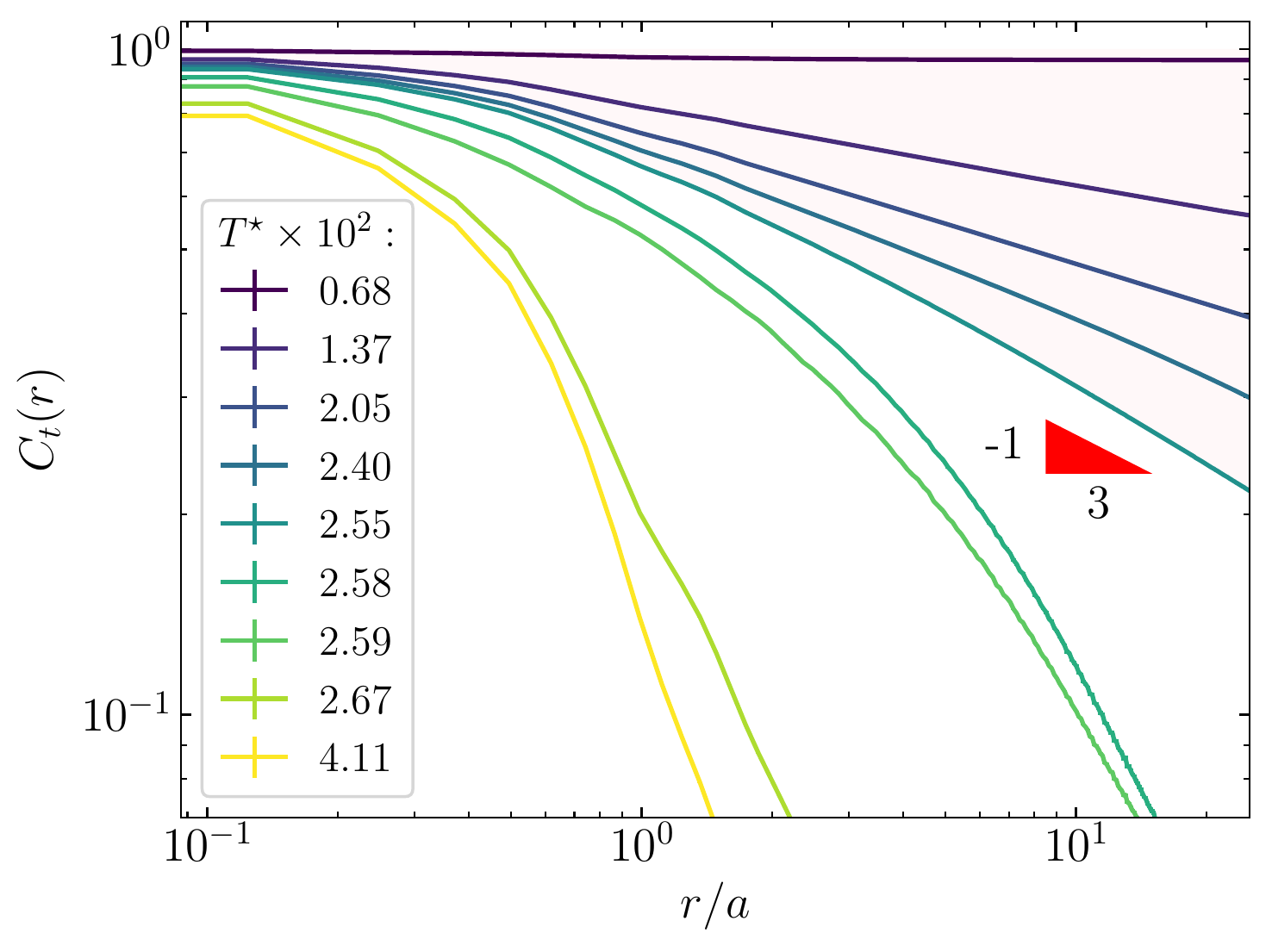}
		\label{fig:plots_corrT_lin}
	}
	\subfloat[]{
		\includegraphics[width=0.5\columnwidth]{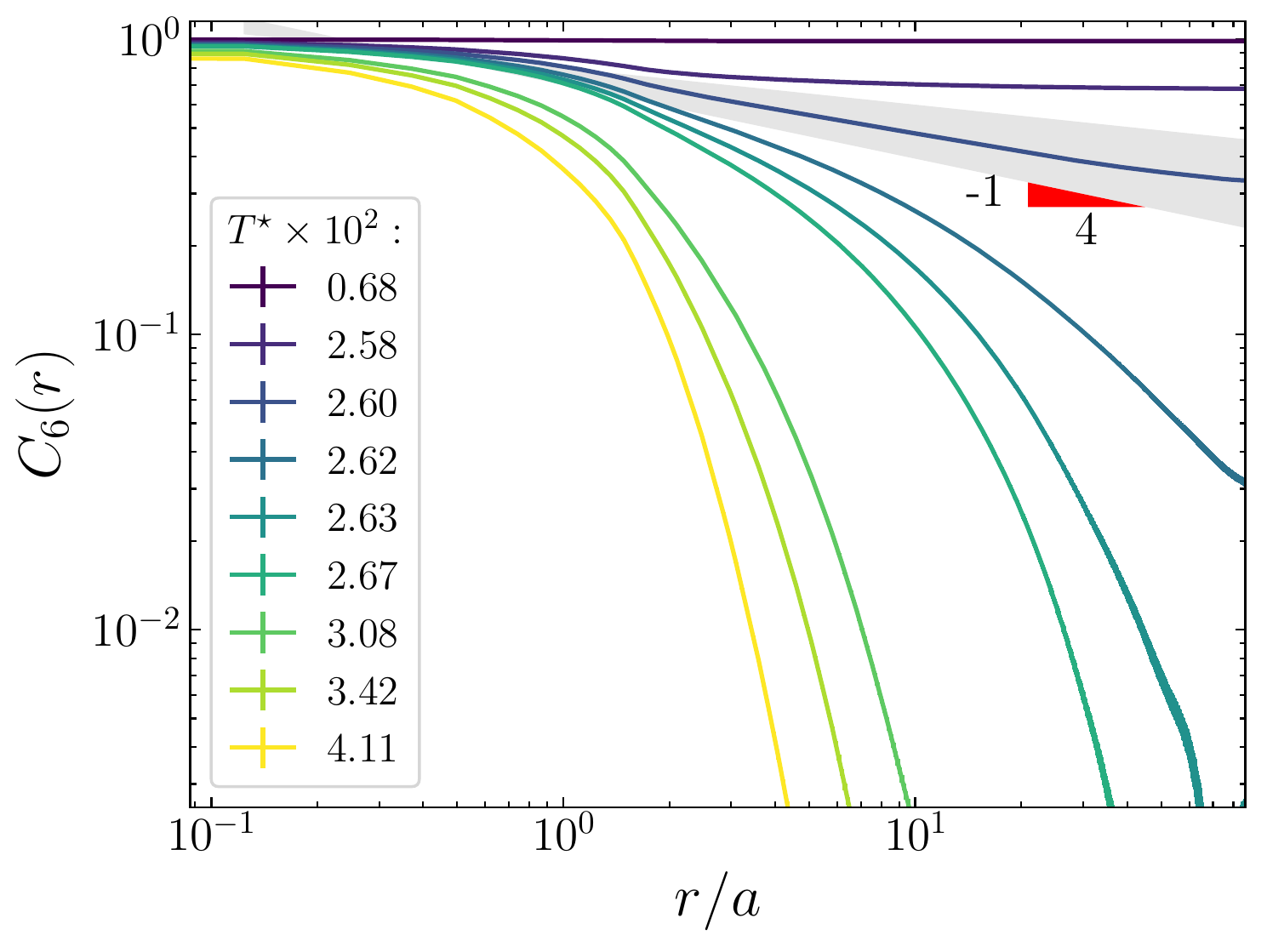}
		\label{fig:plots_corr6_lin}
	}		
	\caption{(a) Translational correlation function $C_t(r)$ in a log-log plot. The curve \revision{evolves} from a power-law decay (pink area) to an exponential decay as temperature increases. The $r^{-1/3}$ decaying behaviour at the solid-hexatic transition is predicted by KTHNY theory. (b) Orientational correlation function $C_6(r)$ in a log-log plot. At low temperatures, the curves decay to a plateau. At intermediate temperatures, curves follow a power-law decay (gray area). At higher temperatures, the curves exhibit an exponential decays. The $r^{-1/4}$ decaying behaviour at the hexatic-liquid transition is predicted by KTHNY theory.}
\end{figure}
In
Fig. \ref{fig:plots_corrT_lin}, $C_t(r)$ shows two different decaying behaviors: for $T^\star \lesssim 2.55\times 10^{-2}$, $C_t(r)$ decays algebraically, which is a signature of a quasi-long range positional order, and is typical of a 2D solid phase. For $T^\star \gtrsim 2.55\times 10^{-2}$, $C_t(r)$ decays exponentially, revealing a short range positional order. Near the transition between the two regimes, the power-law exponent is close to $-1/3$, in agreement with the prediction of the KTHNY theory.

In
Fig. \ref{fig:plots_corr6_lin}, $C_6(r)$ exhibits three different behaviors:
at low temperatures ($T^\star \lesssim 2.55\times 10^{-2}$), $C_6(r)$ approaches constants, and the
system is in the solid phase with long-range orientational
order.
At intermediate temperature ($2.55\times 10^{-2} \lesssim T^\star \lesssim 2.61\times 10^{-2}$), $C_6(r)$ decays algebraically with an exponent close to $-1/4$, which agrees with the prediction of the KTHNY
theory and confirms the existence of an hexatic phase.  When the temperature is further increased, $C_6(r)$ decays exponentially and the system becomes a liquid.

\begin{figure}[h]
	\centering
	\subfloat[]{	
	\includegraphics[width=0.75\columnwidth]{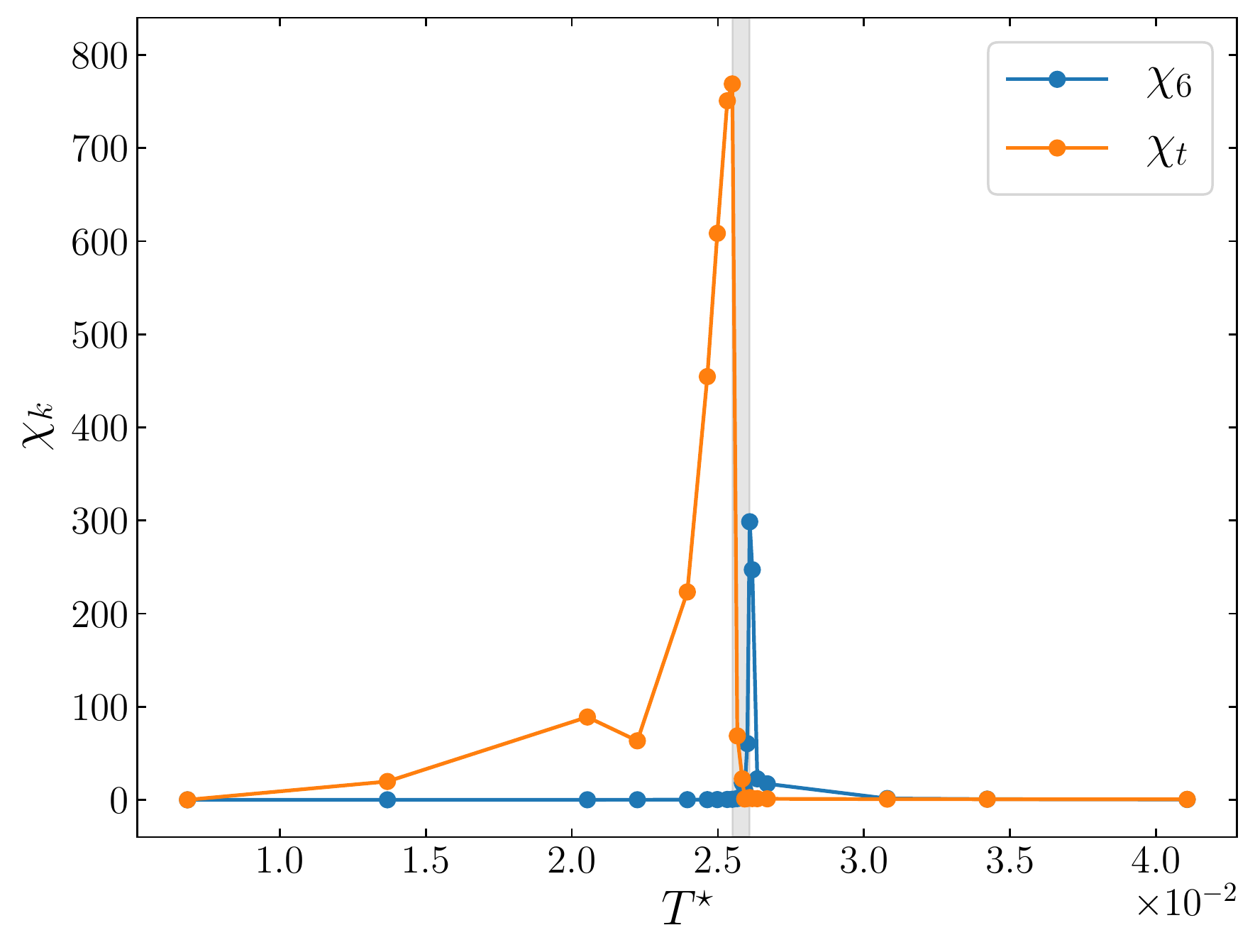}
	\label{fig:plot_susceptibilities}
	}

	\caption{The susceptibilities, $\chi_t$ and $\chi_6$ as a function of reduced temperature. The peaks of the translational susceptibility $\chi_t$ and the orientational susceptibility $\chi_6$ clearly indicate two transition points, at $T^\star_{m}=2.549\times 10^{-2}$ and $T^\star_{i}=2.609\times 10^{-2}$ respectively.
	}
\end{figure}

To determine more precisely the phase-transition points, we plot the susceptibilities associated with the two order parameters. Unlike correlation functions, the divergence of susceptibility has been shown
to be robust to finite-size or finite-time effects.
%
Fig. \ref{fig:plot_susceptibilities} shows the variation of the \revision{susceptibilities 
$\chi_k=N\left( \langle \vert \Psi_k \vert^2 \rangle - \langle \vert \Psi_k \vert \rangle^2 \right)$ with temperature.
}

Their sharp peaks clearly indicate two transitions in the melting process. The peak for $\chi_t$ is centered at $T^\star_{m}=2.549\times 10^{-2}$, while the peak for $\chi_6$ is centered at $T^\star_{i}=2.609\times 10^{-2}$, confirming the existence of an intermediate hexatic phase. It must be emphasized that the values of $T^\star_{m}$, $T^\star_{i}$ we find are specific to the value of the reduced compressibilitiy $\xi$ chosen for our simulations.
The peak for $\chi_t$ is much higher than for $\chi_6$,
suggesting a continuous solid-hexatic transition and a first-order hexatic-liquid transition. This modified KTHNY scenario has also been observerd for systems of hard disks \cite{Krauth_2011}.

\begin{figure}[h]
	\centering
	\includegraphics[width=0.75\columnwidth]{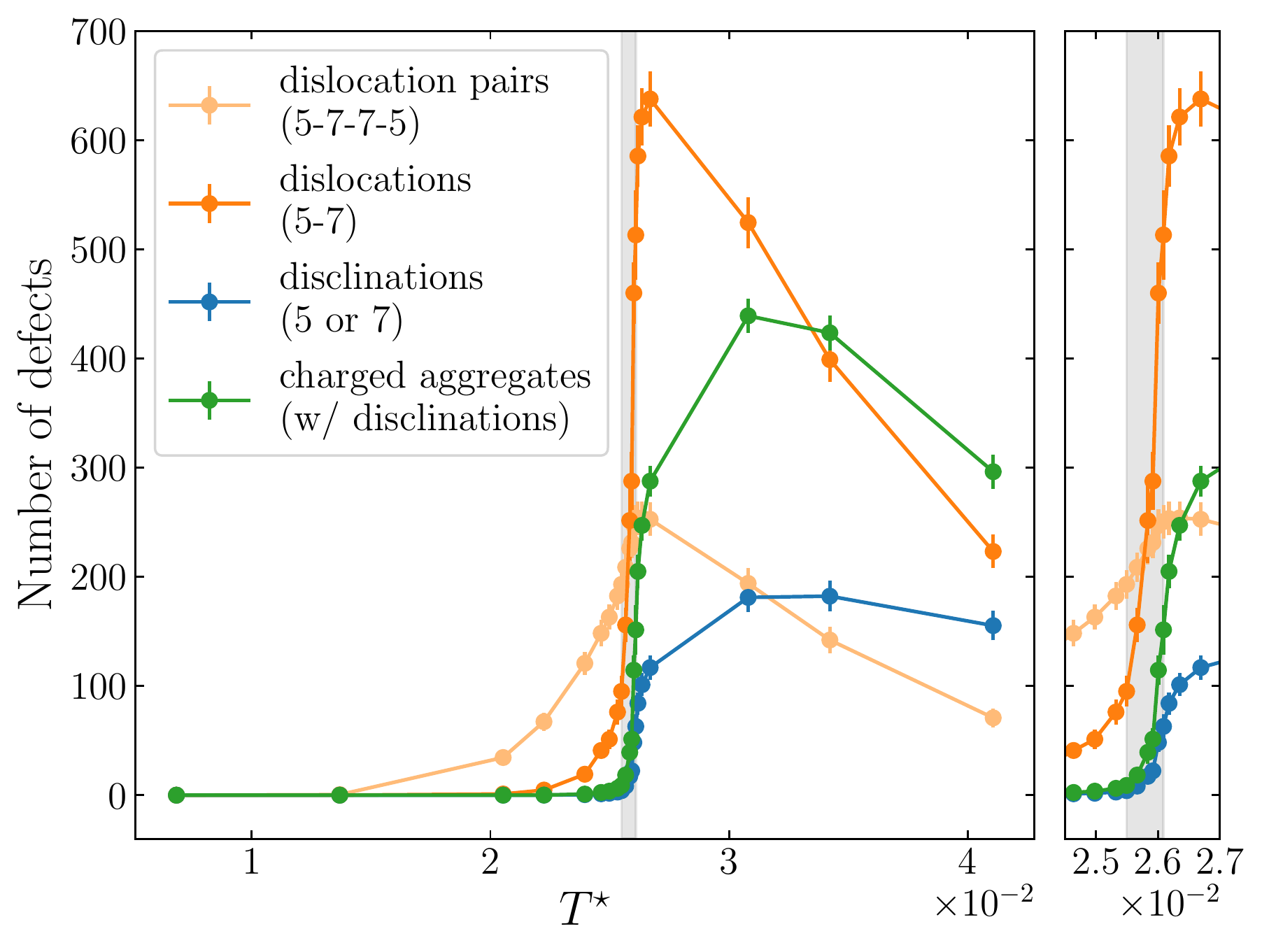}
	\caption{Number of bound dislocation pairs, dislocations, disclinations, and charged aggregates as a function of reduced temperature. Right panel: close-up view showing that the two phase transitions coincide with the rapid increase of dislocations and disclinations+charged aggregates, respectively.}
	\label{fig:defects_vs_T}
\end{figure}

KTHNY theory relates the disappearance of translational and rotational orders to dislocations and disclinations unbinding, respectively. To test this scenario, we report on Fig. \ref{fig:defects_vs_T} the evolution of population of defects with temperature. \revision{Detection and counting of the different defects are detailed in the Supplemental Material \cite{Suppl_Mat}.} As expected, $T^\star_{m}$ coincides with the rapid increase of isolated dislocations due to the the dissociation of bound dislocation pairs. On the other hand, the number of disclinations shows a moderate increase around $T^\star_{i}$. 
In fact, although the KTHNY scenario assumes that the \revision{defects remain} diluted during the melting process, in our systems the concentration of defects is such that they form aggregates whose number and mean size increase with temperature. This aggregation is a natural consequence of the attractive interactions between defects. Aggregates with a non-zero topological charge (\ie containing unequal numbers of 5-sided and 7-sided cells) participate as much as free disclinations in the destruction of the orientational order, and hence must be accounted for. Fig. \ref{fig:defects_vs_T} shows that the population of charged aggregates (including disclinations) increases significantly around $T^\star_{i}$, in good agreement with the KTHNY scenario.

In summary, we used a recent
CPM modified algorithm that allows for thermalization of large cellular systems, and showed that soft cellular systems follow closely the KTHNY melting scenario, hence extending the validity of this theory to systems with many-body interactions. We showed in particular the existence of an intermediate hexatic phase. 
Topological properties of SCS have been mainly characterized by $p(n)$, the proportion of $n$-sided cells within the system (and many times by the second moment of this distribution solely). Our study shows that $p(n)$ is often not sufficient to capture the mechanical properties of the system, as it can be in a solid, hexatic, or liquid phase.  Spatial correlations of defects must also be accounted for. We hope
our results will stimulate relevant experimental work to test
the existence of an intermediate phase in the order-disorder transitions observed during morphogenetic movements \cite{Zallen_2004,Classen_2005,Hocevar_2009}.
As the defect core
energy is the vital predictor of the melting mechanism between KTHNY and grain-boundary scenarios, it would be valuable to quantify the defect core energy of SCS in the future.

\bibliography{biblio}

\begin{thebibliography}{40}
\expandafter\ifx\csname natexlab\endcsname\relax\def\natexlab#1{#1}\fi
\expandafter\ifx\csname bibnamefont\endcsname\relax
  \def\bibnamefont#1{#1}\fi
\expandafter\ifx\csname bibfnamefont\endcsname\relax
  \def\bibfnamefont#1{#1}\fi
\expandafter\ifx\csname citenamefont\endcsname\relax
  \def\citenamefont#1{#1}\fi
\expandafter\ifx\csname url\endcsname\relax
  \def\url#1{\texttt{#1}}\fi
\expandafter\ifx\csname urlprefix\endcsname\relax\def\urlprefix{URL }\fi
\providecommand{\bibinfo}[2]{#2}
\providecommand{\eprint}[2][]{\url{#2}}

\bibitem[{Foo()}]{Footnote}
\bibinfo{note}{Other structural rearrangements which do not preserve the number
  and sizes of of the cellular units -- such as cell division or apoptosis in
  biological tissues, or coarsening in foams and emulsions -- usually occur on
  longer time scales.}

\bibitem[{\citenamefont{Aste and Sherrington}(1999)}]{Aste_1999}
\bibinfo{author}{\bibfnamefont{T.}~\bibnamefont{Aste}} \bibnamefont{and}
  \bibinfo{author}{\bibfnamefont{D.}~\bibnamefont{Sherrington}},
  \bibinfo{journal}{Journal of Physics A: Mathematical and General}
  \textbf{\bibinfo{volume}{32}}, \bibinfo{pages}{7049} (\bibinfo{year}{1999}).

\bibitem[{\citenamefont{Davison and Sherrington}(2000)}]{Davison_2000}
\bibinfo{author}{\bibfnamefont{L.}~\bibnamefont{Davison}} \bibnamefont{and}
  \bibinfo{author}{\bibfnamefont{D.}~\bibnamefont{Sherrington}},
  \bibinfo{journal}{Journal of Physics A: Mathematical and General}
  \textbf{\bibinfo{volume}{33}}, \bibinfo{pages}{8615} (\bibinfo{year}{2000}).

\bibitem[{\citenamefont{Angelini et~al.}(2011)\citenamefont{Angelini, Hannezo,
  Trepat, Marquez, Fredberg, and Weitz}}]{Angelini_2011}
\bibinfo{author}{\bibfnamefont{T.~E.} \bibnamefont{Angelini}},
  \bibinfo{author}{\bibfnamefont{E.}~\bibnamefont{Hannezo}},
  \bibinfo{author}{\bibfnamefont{X.}~\bibnamefont{Trepat}},
  \bibinfo{author}{\bibfnamefont{M.}~\bibnamefont{Marquez}},
  \bibinfo{author}{\bibfnamefont{J.~J.} \bibnamefont{Fredberg}},
  \bibnamefont{and} \bibinfo{author}{\bibfnamefont{D.~A.} \bibnamefont{Weitz}},
  \bibinfo{journal}{Proceedings of the National Academy of Sciences}
  \textbf{\bibinfo{volume}{108}}, \bibinfo{pages}{4714} (\bibinfo{year}{2011}).

\bibitem[{\citenamefont{Bi et~al.}(2015)\citenamefont{Bi, Lopez, Schwarz, and
  Manning}}]{Bi_2015}
\bibinfo{author}{\bibfnamefont{D.}~\bibnamefont{Bi}},
  \bibinfo{author}{\bibfnamefont{J.~H.} \bibnamefont{Lopez}},
  \bibinfo{author}{\bibfnamefont{J.~M.} \bibnamefont{Schwarz}},
  \bibnamefont{and} \bibinfo{author}{\bibfnamefont{M.~L.}
  \bibnamefont{Manning}}, \bibinfo{journal}{Nature Physics}
  \textbf{\bibinfo{volume}{11}}, \bibinfo{pages}{1074} (\bibinfo{year}{2015}),
  ISSN \bibinfo{issn}{1745-2473, 1745-2481}.

\bibitem[{\citenamefont{Bi et~al.}(2016)\citenamefont{Bi, Yang, Marchetti, and
  Manning}}]{Bi_2016}
\bibinfo{author}{\bibfnamefont{D.}~\bibnamefont{Bi}},
  \bibinfo{author}{\bibfnamefont{X.}~\bibnamefont{Yang}},
  \bibinfo{author}{\bibfnamefont{M.~C.} \bibnamefont{Marchetti}},
  \bibnamefont{and} \bibinfo{author}{\bibfnamefont{M.~L.}
  \bibnamefont{Manning}}, \bibinfo{journal}{Physical Review X}
  \textbf{\bibinfo{volume}{6}} (\bibinfo{year}{2016}), ISSN
  \bibinfo{issn}{2160-3308}.

\bibitem[{\citenamefont{Quilliet et~al.}(2008)\citenamefont{Quilliet, Talebi,
  Rabaud, Käfer, Cox, and Graner}}]{Quilliet_2008}
\bibinfo{author}{\bibfnamefont{C.}~\bibnamefont{Quilliet}},
  \bibinfo{author}{\bibfnamefont{S.~A.} \bibnamefont{Talebi}},
  \bibinfo{author}{\bibfnamefont{D.}~\bibnamefont{Rabaud}},
  \bibinfo{author}{\bibfnamefont{J.}~\bibnamefont{Käfer}},
  \bibinfo{author}{\bibfnamefont{S.}~\bibnamefont{Cox}}, \bibnamefont{and}
  \bibinfo{author}{\bibfnamefont{F.}~\bibnamefont{Graner}},
  \bibinfo{journal}{Philosophical Magazine Letters}
  \textbf{\bibinfo{volume}{88}}, \bibinfo{pages}{651} (\bibinfo{year}{2008}).

\bibitem[{\citenamefont{Durand et~al.}(2011)\citenamefont{Durand, Käfer,
  Quilliet, Cox, Talebi, and Graner}}]{Durand_statistical_2011}
\bibinfo{author}{\bibfnamefont{M.}~\bibnamefont{Durand}},
  \bibinfo{author}{\bibfnamefont{J.}~\bibnamefont{Käfer}},
  \bibinfo{author}{\bibfnamefont{C.}~\bibnamefont{Quilliet}},
  \bibinfo{author}{\bibfnamefont{S.}~\bibnamefont{Cox}},
  \bibinfo{author}{\bibfnamefont{S.~A.} \bibnamefont{Talebi}},
  \bibnamefont{and} \bibinfo{author}{\bibfnamefont{F.}~\bibnamefont{Graner}},
  \bibinfo{journal}{Physical Review Letters} \textbf{\bibinfo{volume}{107}}
  (\bibinfo{year}{2011}), ISSN \bibinfo{issn}{0031-9007, 1079-7114}.

\bibitem[{\citenamefont{Durand et~al.}(2014)\citenamefont{Durand, Kraynik, van
  Swol, Käfer, Quilliet, Cox, Ataei~Talebi, and
  Graner}}]{Durand_statistical_2014}
\bibinfo{author}{\bibfnamefont{M.}~\bibnamefont{Durand}},
  \bibinfo{author}{\bibfnamefont{A.~M.} \bibnamefont{Kraynik}},
  \bibinfo{author}{\bibfnamefont{F.}~\bibnamefont{van Swol}},
  \bibinfo{author}{\bibfnamefont{J.}~\bibnamefont{Käfer}},
  \bibinfo{author}{\bibfnamefont{C.}~\bibnamefont{Quilliet}},
  \bibinfo{author}{\bibfnamefont{S.}~\bibnamefont{Cox}},
  \bibinfo{author}{\bibfnamefont{S.}~\bibnamefont{Ataei~Talebi}},
  \bibnamefont{and} \bibinfo{author}{\bibfnamefont{F.}~\bibnamefont{Graner}},
  \bibinfo{journal}{Physical Review E} \textbf{\bibinfo{volume}{89}}
  (\bibinfo{year}{2014}), ISSN \bibinfo{issn}{1539-3755, 1550-2376}.

\bibitem[{\citenamefont{Zallen and Zallen}(2004)}]{Zallen_2004}
\bibinfo{author}{\bibfnamefont{J.~A.} \bibnamefont{Zallen}} \bibnamefont{and}
  \bibinfo{author}{\bibfnamefont{R.}~\bibnamefont{Zallen}},
  \bibinfo{journal}{Journal of Physics: Condensed Matter}
  \textbf{\bibinfo{volume}{16}}, \bibinfo{pages}{S5073} (\bibinfo{year}{2004}).

\bibitem[{\citenamefont{Classen et~al.}(2005)\citenamefont{Classen, Anderson,
  Marois, and Eaton}}]{Classen_2005}
\bibinfo{author}{\bibfnamefont{A.-K.} \bibnamefont{Classen}},
  \bibinfo{author}{\bibfnamefont{K.~I.} \bibnamefont{Anderson}},
  \bibinfo{author}{\bibfnamefont{E.}~\bibnamefont{Marois}}, \bibnamefont{and}
  \bibinfo{author}{\bibfnamefont{S.}~\bibnamefont{Eaton}},
  \bibinfo{journal}{Developmental Cell} \textbf{\bibinfo{volume}{9}},
  \bibinfo{pages}{805 } (\bibinfo{year}{2005}), ISSN \bibinfo{issn}{1534-5807}.

\bibitem[{\citenamefont{Hočevar and Ziherl}(2009)}]{Hocevar_2009}
\bibinfo{author}{\bibfnamefont{A.}~\bibnamefont{Hočevar}} \bibnamefont{and}
  \bibinfo{author}{\bibfnamefont{P.}~\bibnamefont{Ziherl}},
  \bibinfo{journal}{Physical Review E} \textbf{\bibinfo{volume}{80}}
  (\bibinfo{year}{2009}), ISSN \bibinfo{issn}{1539-3755, 1550-2376}.

\bibitem[{\citenamefont{Staple et~al.}(2010)\citenamefont{Staple, Farhadifar,
  Röper, Aigouy, Eaton, and Jülicher}}]{Staple_2010}
\bibinfo{author}{\bibfnamefont{D.~B.} \bibnamefont{Staple}},
  \bibinfo{author}{\bibfnamefont{R.}~\bibnamefont{Farhadifar}},
  \bibinfo{author}{\bibfnamefont{J.~C.} \bibnamefont{Röper}},
  \bibinfo{author}{\bibfnamefont{B.}~\bibnamefont{Aigouy}},
  \bibinfo{author}{\bibfnamefont{S.}~\bibnamefont{Eaton}}, \bibnamefont{and}
  \bibinfo{author}{\bibfnamefont{F.}~\bibnamefont{Jülicher}},
  \bibinfo{journal}{The European Physical Journal E}
  \textbf{\bibinfo{volume}{33}}, \bibinfo{pages}{117} (\bibinfo{year}{2010}),
  ISSN \bibinfo{issn}{1292-895X}.

\bibitem[{\citenamefont{Sussman et~al.}(2018)\citenamefont{Sussman, Schwarz,
  Marchetti, and Manning}}]{Sussman_2018}
\bibinfo{author}{\bibfnamefont{D.~M.} \bibnamefont{Sussman}},
  \bibinfo{author}{\bibfnamefont{J.}~\bibnamefont{Schwarz}},
  \bibinfo{author}{\bibfnamefont{M.~C.} \bibnamefont{Marchetti}},
  \bibnamefont{and} \bibinfo{author}{\bibfnamefont{M.~L.}
  \bibnamefont{Manning}}, \bibinfo{journal}{Physical Review Letters}
  \textbf{\bibinfo{volume}{120}} (\bibinfo{year}{2018}), ISSN
  \bibinfo{issn}{0031-9007, 1079-7114}.

\bibitem[{\citenamefont{Fodor et~al.}(2018)\citenamefont{Fodor, Mehandia,
  Comelles, Thiagarajan, Gov, Visco, van Wijland, and Riveline}}]{Fodor_2018}
\bibinfo{author}{\bibfnamefont{E.}~\bibnamefont{Fodor}},
  \bibinfo{author}{\bibfnamefont{V.}~\bibnamefont{Mehandia}},
  \bibinfo{author}{\bibfnamefont{J.}~\bibnamefont{Comelles}},
  \bibinfo{author}{\bibfnamefont{R.}~\bibnamefont{Thiagarajan}},
  \bibinfo{author}{\bibfnamefont{N.~S.} \bibnamefont{Gov}},
  \bibinfo{author}{\bibfnamefont{P.}~\bibnamefont{Visco}},
  \bibinfo{author}{\bibfnamefont{F.}~\bibnamefont{van Wijland}},
  \bibnamefont{and} \bibinfo{author}{\bibfnamefont{D.}~\bibnamefont{Riveline}},
  \bibinfo{journal}{Biophysical Journal} \textbf{\bibinfo{volume}{114}},
  \bibinfo{pages}{939} (\bibinfo{year}{2018}), ISSN \bibinfo{issn}{00063495}.

\bibitem[{\citenamefont{Bragg and Nye}(1947)}]{Bragg_1947}
\bibinfo{author}{\bibfnamefont{W.~L.} \bibnamefont{Bragg}} \bibnamefont{and}
  \bibinfo{author}{\bibfnamefont{J.~F.} \bibnamefont{Nye}},
  \bibinfo{journal}{Proceedings of the Royal Society of London A: Mathematical,
  Physical and Engineering Sciences} \textbf{\bibinfo{volume}{190}},
  \bibinfo{pages}{474} (\bibinfo{year}{1947}), ISSN \bibinfo{issn}{0080-4630}.

\bibitem[{\citenamefont{Bragg and Lomer}(1949)}]{Bragg_1949}
\bibinfo{author}{\bibfnamefont{W.~L.} \bibnamefont{Bragg}} \bibnamefont{and}
  \bibinfo{author}{\bibfnamefont{W.~M.} \bibnamefont{Lomer}},
  \bibinfo{journal}{Proceedings of the Royal Society of London A: Mathematical,
  Physical and Engineering Sciences} \textbf{\bibinfo{volume}{196}},
  \bibinfo{pages}{171} (\bibinfo{year}{1949}), ISSN \bibinfo{issn}{0080-4630}.

\bibitem[{\citenamefont{Chen et~al.}(1995)\citenamefont{Chen, Kaplan, and
  Mostoller}}]{Chen_1995}
\bibinfo{author}{\bibfnamefont{K.}~\bibnamefont{Chen}},
  \bibinfo{author}{\bibfnamefont{T.}~\bibnamefont{Kaplan}}, \bibnamefont{and}
  \bibinfo{author}{\bibfnamefont{M.}~\bibnamefont{Mostoller}},
  \bibinfo{journal}{Phys. Rev. Lett.} \textbf{\bibinfo{volume}{74}},
  \bibinfo{pages}{4019} (\bibinfo{year}{1995}).

\bibitem[{\citenamefont{von Grünberg et~al.}(2007)\citenamefont{von Grünberg,
  Keim, and Maret}}]{vonGrunberg_2007}
\bibinfo{author}{\bibfnamefont{H.~H.} \bibnamefont{von Grünberg}},
  \bibinfo{author}{\bibfnamefont{P.}~\bibnamefont{Keim}}, \bibnamefont{and}
  \bibinfo{author}{\bibfnamefont{G.}~\bibnamefont{Maret}}, in
  \emph{\bibinfo{booktitle}{Colloidal order : entropic and surface forces}},
  edited by \bibinfo{editor}{\bibfnamefont{G.}~\bibnamefont{Gompper}}
  (\bibinfo{publisher}{WILEY-VCH}, \bibinfo{address}{Weinheim},
  \bibinfo{year}{2007}), no.~\bibinfo{number}{3} in \bibinfo{series}{Soft
  matter}, pp. \bibinfo{pages}{40--83}, ISBN \bibinfo{isbn}{978-3-527-31370-9}.

\bibitem[{\citenamefont{Marcus and Rice}(1996)}]{Marcus_1996}
\bibinfo{author}{\bibfnamefont{A.~H.} \bibnamefont{Marcus}} \bibnamefont{and}
  \bibinfo{author}{\bibfnamefont{S.~A.} \bibnamefont{Rice}},
  \bibinfo{journal}{Phys. Rev. Lett.} \textbf{\bibinfo{volume}{77}},
  \bibinfo{pages}{2577} (\bibinfo{year}{1996}).

\bibitem[{\citenamefont{Schockmel et~al.}(2013)\citenamefont{Schockmel, Mersch,
  Vandewalle, and Lumay}}]{Schockmel_2013}
\bibinfo{author}{\bibfnamefont{J.}~\bibnamefont{Schockmel}},
  \bibinfo{author}{\bibfnamefont{E.}~\bibnamefont{Mersch}},
  \bibinfo{author}{\bibfnamefont{N.}~\bibnamefont{Vandewalle}},
  \bibnamefont{and} \bibinfo{author}{\bibfnamefont{G.}~\bibnamefont{Lumay}},
  \bibinfo{journal}{Physical Review E} \textbf{\bibinfo{volume}{87}}
  (\bibinfo{year}{2013}), ISSN \bibinfo{issn}{1539-3755, 1550-2376}.

\bibitem[{\citenamefont{Kapfer and Krauth}(2015)}]{Kapfer_2015}
\bibinfo{author}{\bibfnamefont{S.~C.} \bibnamefont{Kapfer}} \bibnamefont{and}
  \bibinfo{author}{\bibfnamefont{W.}~\bibnamefont{Krauth}},
  \bibinfo{journal}{Phys. Rev. Lett.} \textbf{\bibinfo{volume}{114}},
  \bibinfo{pages}{035702} (\bibinfo{year}{2015}).

\bibitem[{\citenamefont{Höhler and Cohen-Addad}(2017)}]{Hohler_2017}
\bibinfo{author}{\bibfnamefont{R.}~\bibnamefont{Höhler}} \bibnamefont{and}
  \bibinfo{author}{\bibfnamefont{S.}~\bibnamefont{Cohen-Addad}},
  \bibinfo{journal}{Soft Matter} \textbf{\bibinfo{volume}{13}},
  \bibinfo{pages}{1371} (\bibinfo{year}{2017}).

\bibitem[{\citenamefont{Ginot et~al.}(2019)\citenamefont{Ginot, Höhler,
  Mariot, Kraynik, and Drenckhan}}]{Ginot_2019}
\bibinfo{author}{\bibfnamefont{G.}~\bibnamefont{Ginot}},
  \bibinfo{author}{\bibfnamefont{R.}~\bibnamefont{Höhler}},
  \bibinfo{author}{\bibfnamefont{S.}~\bibnamefont{Mariot}},
  \bibinfo{author}{\bibfnamefont{A.}~\bibnamefont{Kraynik}}, \bibnamefont{and}
  \bibinfo{author}{\bibfnamefont{W.}~\bibnamefont{Drenckhan}},
  \bibinfo{journal}{Soft Matter} \textbf{\bibinfo{volume}{15}},
  \bibinfo{pages}{4570} (\bibinfo{year}{2019}).

\bibitem[{\citenamefont{Strandburg}(1988)}]{Strandburg_1988}
\bibinfo{author}{\bibfnamefont{K.~J.} \bibnamefont{Strandburg}},
  \bibinfo{journal}{Reviews of modern physics} \textbf{\bibinfo{volume}{60}},
  \bibinfo{pages}{161} (\bibinfo{year}{1988}).

\bibitem[{\citenamefont{Glaser and Clark}(1993)}]{Glaser_1993}
\bibinfo{author}{\bibfnamefont{M.~A.} \bibnamefont{Glaser}} \bibnamefont{and}
  \bibinfo{author}{\bibfnamefont{N.~A.} \bibnamefont{Clark}},
  \emph{\bibinfo{title}{Melting and Liquid Structure in two Dimensions}}
  (\bibinfo{publisher}{John Wiley \& Sons, Inc.}, \bibinfo{year}{1993}), pp.
  \bibinfo{pages}{543--709}, ISBN \bibinfo{isbn}{9780470141410}.

\bibitem[{\citenamefont{Saito}(1982)}]{Saito_1982}
\bibinfo{author}{\bibfnamefont{Y.}~\bibnamefont{Saito}},
  \bibinfo{journal}{Physical Review B} \textbf{\bibinfo{volume}{26}},
  \bibinfo{pages}{6239} (\bibinfo{year}{1982}).

\bibitem[{\citenamefont{Chui}(1983)}]{Chui_1983}
\bibinfo{author}{\bibfnamefont{S.~T.} \bibnamefont{Chui}},
  \bibinfo{journal}{Physical Review B} \textbf{\bibinfo{volume}{28}},
  \bibinfo{pages}{178} (\bibinfo{year}{1983}).

\bibitem[{\citenamefont{Glazier et~al.}(1990)\citenamefont{Glazier, Anderson,
  and Grest}}]{Glazier1990}
\bibinfo{author}{\bibfnamefont{J.~A.} \bibnamefont{Glazier}},
  \bibinfo{author}{\bibfnamefont{M.~P.} \bibnamefont{Anderson}},
  \bibnamefont{and} \bibinfo{author}{\bibfnamefont{G.~S.} \bibnamefont{Grest}},
  \bibinfo{journal}{Philos. Mag. B} \textbf{\bibinfo{volume}{62}},
  \bibinfo{pages}{615} (\bibinfo{year}{1990}).

\bibitem[{\citenamefont{Jiang et~al.}(1999)\citenamefont{Jiang, Swart, Saxena,
  Asipauskas, and Glazier}}]{Jiang_1999}
\bibinfo{author}{\bibfnamefont{Y.}~\bibnamefont{Jiang}},
  \bibinfo{author}{\bibfnamefont{P.~J.} \bibnamefont{Swart}},
  \bibinfo{author}{\bibfnamefont{A.}~\bibnamefont{Saxena}},
  \bibinfo{author}{\bibfnamefont{M.}~\bibnamefont{Asipauskas}},
  \bibnamefont{and} \bibinfo{author}{\bibfnamefont{J.~A.}
  \bibnamefont{Glazier}}, \bibinfo{journal}{Physical Review E}
  \textbf{\bibinfo{volume}{59}}, \bibinfo{pages}{5819} (\bibinfo{year}{1999}).

\bibitem[{\citenamefont{Hirashima et~al.}(2017)\citenamefont{Hirashima, Rens,
  and Merks}}]{Hirashima_2017}
\bibinfo{author}{\bibfnamefont{T.}~\bibnamefont{Hirashima}},
  \bibinfo{author}{\bibfnamefont{E.~G.} \bibnamefont{Rens}}, \bibnamefont{and}
  \bibinfo{author}{\bibfnamefont{R.~M.~H.} \bibnamefont{Merks}},
  \bibinfo{journal}{Development, Growth \& Differentiation}
  \textbf{\bibinfo{volume}{59}}, \bibinfo{pages}{329} (\bibinfo{year}{2017}).

\bibitem[{\citenamefont{Graner and Glazier}(1992)}]{Graner_1992}
\bibinfo{author}{\bibfnamefont{F.}~\bibnamefont{Graner}} \bibnamefont{and}
  \bibinfo{author}{\bibfnamefont{J.~A.} \bibnamefont{Glazier}},
  \bibinfo{journal}{Physical Review Letters} \textbf{\bibinfo{volume}{69}},
  \bibinfo{pages}{2013} (\bibinfo{year}{1992}).

\bibitem[{\citenamefont{Szab{\'{o}} et~al.}(2010)\citenamefont{Szab{\'{o}},
  \"{U}nnep, M{\'{e}}hes, Twal, Argraves, Cao, and Czir{\'{o}}k}}]{Szabo_2010}
\bibinfo{author}{\bibfnamefont{A.}~\bibnamefont{Szab{\'{o}}}},
  \bibinfo{author}{\bibfnamefont{R.}~\bibnamefont{\"{U}nnep}},
  \bibinfo{author}{\bibfnamefont{E.}~\bibnamefont{M{\'{e}}hes}},
  \bibinfo{author}{\bibfnamefont{W.~O.} \bibnamefont{Twal}},
  \bibinfo{author}{\bibfnamefont{W.~S.} \bibnamefont{Argraves}},
  \bibinfo{author}{\bibfnamefont{Y.}~\bibnamefont{Cao}}, \bibnamefont{and}
  \bibinfo{author}{\bibfnamefont{A.}~\bibnamefont{Czir{\'{o}}k}},
  \bibinfo{journal}{Physical Biology} \textbf{\bibinfo{volume}{7}},
  \bibinfo{pages}{046007} (\bibinfo{year}{2010}).

\bibitem[{\citenamefont{Kabla}(2012)}]{Kabla_2012}
\bibinfo{author}{\bibfnamefont{A.~J.} \bibnamefont{Kabla}},
  \bibinfo{journal}{Journal of The Royal Society Interface} p.
  \bibinfo{pages}{rsif20120448} (\bibinfo{year}{2012}).

\bibitem[{\citenamefont{Durand and Guesnet}(2016)}]{Durand_2016}
\bibinfo{author}{\bibfnamefont{M.}~\bibnamefont{Durand}} \bibnamefont{and}
  \bibinfo{author}{\bibfnamefont{E.}~\bibnamefont{Guesnet}},
  \bibinfo{journal}{Computer Physics Communications}
  \textbf{\bibinfo{volume}{208}}, \bibinfo{pages}{54} (\bibinfo{year}{2016}),
  ISSN \bibinfo{issn}{00104655}.

\bibitem[{Sup()}]{Suppl_Mat}
\bibinfo{note}{See Supplemental Material at [URL will be inserted by publisher]
  for further details on numerical simulations, which includes Refs.
  \cite{Ouchi_2003,Magno_2015, Farhadifar_2007, Bi_2015}.}

\bibitem[{\citenamefont{Bernard and Krauth}(2011)}]{Krauth_2011}
\bibinfo{author}{\bibfnamefont{E.~P.} \bibnamefont{Bernard}} \bibnamefont{and}
  \bibinfo{author}{\bibfnamefont{W.}~\bibnamefont{Krauth}},
  \bibinfo{journal}{Physical Review Letters} \textbf{\bibinfo{volume}{107}}
  (\bibinfo{year}{2011}), ISSN \bibinfo{issn}{0031-9007, 1079-7114}.

\bibitem[{\citenamefont{Ouchi et~al.}(2003)\citenamefont{Ouchi, Glazier, Rieu,
  Upadhyaya, and Sawada}}]{Ouchi_2003}
\bibinfo{author}{\bibfnamefont{N.~B.} \bibnamefont{Ouchi}},
  \bibinfo{author}{\bibfnamefont{J.~A.} \bibnamefont{Glazier}},
  \bibinfo{author}{\bibfnamefont{J.-P.} \bibnamefont{Rieu}},
  \bibinfo{author}{\bibfnamefont{A.}~\bibnamefont{Upadhyaya}},
  \bibnamefont{and} \bibinfo{author}{\bibfnamefont{Y.}~\bibnamefont{Sawada}},
  \bibinfo{journal}{Physica A: Statistical Mechanics and its Applications}
  \textbf{\bibinfo{volume}{329}}, \bibinfo{pages}{451} (\bibinfo{year}{2003}),
  ISSN \bibinfo{issn}{03784371}.

\bibitem[{\citenamefont{Magno et~al.}(2015)\citenamefont{Magno, Grieneisen, and
  Marée}}]{Magno_2015}
\bibinfo{author}{\bibfnamefont{R.}~\bibnamefont{Magno}},
  \bibinfo{author}{\bibfnamefont{V.~A.} \bibnamefont{Grieneisen}},
  \bibnamefont{and} \bibinfo{author}{\bibfnamefont{A.~F.}
  \bibnamefont{Marée}}, \bibinfo{journal}{BMC Biophysics}
  \textbf{\bibinfo{volume}{8}} (\bibinfo{year}{2015}), ISSN
  \bibinfo{issn}{2046-1682}.

\bibitem[{\citenamefont{Farhadifar et~al.}(2007)\citenamefont{Farhadifar,
  Röper, Aigouy, Eaton, and Jülicher}}]{Farhadifar_2007}
\bibinfo{author}{\bibfnamefont{R.}~\bibnamefont{Farhadifar}},
  \bibinfo{author}{\bibfnamefont{J.-C.} \bibnamefont{Röper}},
  \bibinfo{author}{\bibfnamefont{B.}~\bibnamefont{Aigouy}},
  \bibinfo{author}{\bibfnamefont{S.}~\bibnamefont{Eaton}}, \bibnamefont{and}
  \bibinfo{author}{\bibfnamefont{F.}~\bibnamefont{Jülicher}},
  \bibinfo{journal}{Current Biology} \textbf{\bibinfo{volume}{17}},
  \bibinfo{pages}{2095 } (\bibinfo{year}{2007}), ISSN
  \bibinfo{issn}{0960-9822}.

\end{thebibliography}


\end{document}